\documentstyle[12pt,epsfig]{article}

\newcommand{\ra}{\rightarrow}

\newcommand{\s}{\\ \vspace*{-3mm} }
\newcommand{\nn}{\noindent}
\newcommand{\non}{\nonumber}
\newcommand{\beq}{\begin{eqnarray}}
\newcommand{\eeq}{\end{eqnarray}}

\newcommand{\tb}{\tan\beta}

\newcommand{\ct}[1]{c_{\theta_#1}}
\newcommand{\st}[1]{s_{\theta_#1}}
\newcommand{\sinb}{\sin\beta}
\newcommand{\cosb}{\cos\beta}

\begin{document}

\def\thefootnote{\fnsymbol{footnote}}

\begin{flushright}
KA--TP--20--96\\
September 1996 \\
\end{flushright}

\vspace{1cm}

\begin{center}

{\large\sc {\bf QCD Corrections to Scalar Quark Decays}}

\vspace{1cm}

{\sc A.~Djouadi\footnote{Supported by Deutsche Forschungsgemeinschaft
DFG (Bonn).}, W. Hollik,} and {\sc C. J\"unger}

\vspace{1cm}

Institut f\"ur Theoretische Physik, Universit\"at Karlsruhe,\\
\vspace{0.2cm}
D--76128 Karlsruhe, Germany. 

\end{center}

\vspace{2cm}

\begin{abstract}

\nn In supersymmetric theories, the main decay modes of scalar quarks 
are decays into quarks plus charginos or neutralinos, if the gluinos are 
heavy enough. We calculate the ${\cal O}(\alpha_s)$ QCD corrections 
to these decay modes in the minimal supersymmetric extension of the 
Standard Model. In the case of scalar top and bottom quarks, where mixing 
effects can be important, these corrections can reach values of the 
order of a few ten percent. They can be either positive or negative 
and increase logarithmically with the gluino mass. For the scalar 
partners of light quarks, the corrections do not exceed in general 
the level of ten percent for gluino masses less than 1 TeV.

\end{abstract}

\newpage

\def\thefootnote{\arabic{footnote}}
\setcounter{footnote}{0}

\subsection*{1. Introduction}

Supersymmetric theories (SUSY) \cite{R0,R9} are the best motivated
extensions of the Standard Model (SM) of the electroweak and strong
interactions. They provide an elegant way to stabilize the huge
hierarchy between the Grand Unification or Planck scale and the Fermi 
scale, and its minimal
version, the Minimal Supersymmetric Standard Model (MSSM) allows for a
consistent unification of the gauge coupling constants and a natural
solution of the Dark Matter problem \cite{R1a}. \s

Supersymmetry predicts the existence of a left-- and right--handed scalar
partner to each Standard Model (SM) quark. The current eigenstates,
$\tilde{q}_L$ and $\tilde{q}_R$, mix to give the mass eigenstates
$\tilde{q}_1$ and $\tilde{q}_2$; the mixing angle is proportional to the
quark mass and is therefore important only in the case of the third
generation squarks \cite{R1}. In particular, due to the large value of
the top mass $m_t$, the mixing between the left-- and right--handed
scalar partners of the top quark, $\tilde{t}_L$ and $\tilde{t}_R$, is
very large and after diagonalization of the mass matrix, the lightest
scalar top quark mass eigenstate $\tilde{t}_1$ can be much lighter than the top
quark and all the scalar partners of the light quarks \cite{R1}. \s 

If the gluinos [the spin $1/2$ superpartners of the gluons] are heavy
enough, scalar quarks will mainly decay into quarks and charginos and/or
neutralinos [mixtures of the SUSY partners of the electroweak gauge
bosons and Higgs bosons]. These are in general tree--level two--body
decays, except in the case of the lightest top squark which could decay
into a charm quark and a neutralino through loop diagrams if the decay
into a chargino and a bottom quark is not overwhelming \cite{R2}. These
decays have been extensively discussed in the Born approximation
\cite{R3}. In this paper we will extend these analyses by including the
${\cal O}(\alpha_s)$ corrections, which due to the relatively large
value of the strong coupling constant, might be large and might affect
significantly the decay rates and the branching ratios\footnote{If
the gluinos are lighter than squarks, then squarks will mainly decay
into quarks plus gluinos; the QCD corrections to these processes have
been recently discussed in Refs.\cite{R5,R5a}.}. \s 

The particular case of the QCD corrections to scalar quark decays into
massless quarks and photinos has been discussed in Refs.~\cite{R4,R5}. 
In the general case that we
will address here, there are three [related] features which complicate
the analysis, the common denominator of all these features being the
finite value of quark masses: (i) In the case of the decays of top and bottom
squarks, one needs to take into account the finite value of the top 
quark mass in
the phase space as well as in the loop diagrams. (ii) Scalar quark mixing will
introduce a new parameter which will induce additional contributions;
since the mixing angle appears in the Born approximation, it needs to be
renormalized. (iii) The finite quark mass [which enters the coupling
between scalar quarks, quarks and the neutralino/chargino states] needs 
also to be renormalized. \s
%the regularization of the ultraviolet divergencies needs to
%be performed in a scheme which preserves SUSY and therefore one cannot
%use the dimensional regularization scheme anymore. \s 

The QCD corrections to the reaction $\tilde{q} \ra q \chi$ analyzed in
the present paper are very similar to the case of the reverse process,
$t \ra \tilde{t} \chi^0$ and $t \ra \tilde{b} \chi^+$ recently discussed
in Ref.~\cite{R6} (see also Ref.~\cite{R7}). During the preparation of
this paper, we received a report by Kraml et al. \cite{R8}, where a
similar analysis has been conducted. Our analytical results agree with
those given in this paper\footnote{We thank the Vienna group and in 
particular S. Kraml for their cooperation
in resolving some discrepancies with some of the formulae and plots
given in the early version of the paper Ref.~\cite{R8}. We also thank T.
Plehn for checking independently the results.}. We extend their
numerical analysis, which focused on the decay of the lightest top
squark into the lightest charginos and neutralinos, by discussing the
decays into the heavier charginos and neutralinos and by studying the
case of bottom squarks and the SUSY partners of light squarks. 

\subsection*{2. Born Approximation}

In the Minimal Supersymmetric Standard Model \cite{R0,R9}, there are two
charginos $\chi_i^+ [i=1,2$] and four neutralinos $\chi_{i}^0$
[$i=1$--4]. Their masses and their couplings to squarks and quarks are
given in terms of the Higgs--higgsino mass parameter $\mu$, the ratio of
the vacuum expectation values $\tb$ of the two Higgs doublet MSSM fields
needed to break the electroweak symmetry, and the wino mass parameter
$M_2$. The bino and gluino masses are related to the parameter $M_2$
[$M_1 \sim M_2/2$ and $m_{\tilde{g}} \sim 3.5 M_2$] when the gaugino
masses and the three coupling constants of
SU(3)$\times$SU(2)$\times$U(1) are unified at the Grand Unification
scale. \s 

The squark masses are given in terms of the parameters $\mu$ and $\tb$,
as well as the left-- and right--handed scalar masses $M_{\tilde{q}_L}$
and $M_{\tilde{q}_R}$ [which in general are taken to be equal] and the
soft--SUSY breaking trilinear coupling $A_q$. The top and bottom squark mass
eigenstates, and their mixing angles, are determined by diagonalizing
the following mass matrices 
\begin{equation}
{\cal M}^2_{\tilde{t}} = 
\left( 
  \begin{array}{cc} M_{\tilde{t}_L}^2 + m_t^2 + \cos 2 \beta (\frac{1}{2}
                       - \frac{2}{3}s_W^2) \, M_Z^2  & m_t \, M^{LR}_t \\
                    m_t \, M^{LR}_t & M_{\tilde{t}_R}^2 + m_t^2
                                   + \frac{2}{3}\cos 2 \beta \; s_W^2 \, M_Z^2
  \end{array}
\right)
\end{equation}
\begin{equation}
{\cal M}^2_{\tilde{b}} = 
\left( 
  \begin{array}{cc} M_{\tilde{t}_L}^2 + m_b^2 + \cos 2 \beta (-\frac{1}{2}
                       +\frac{1}{3}s_W^2) \, M_Z^2  & m_b \, M^{LR}_b \\
                    m_b \, M^{LR}_b & M_{\tilde{b}_R}^2 + m_b^2
                                   - \frac{1}{3}\cos 2 \beta \; s_W^2 \, M_Z^2
  \end{array}
\right)
\end{equation}
where $M^{LR}_{t,b}$ in the off--diagonal terms read: $M^{LR}_t = A_t - \mu 
\, \cot \beta$ and $M^{LR}_b = A_b - \mu \tb$. \s

In the Born approximation, the partial widths for the decays 
$\tilde{t}_i \ra t\chi^0_j$, $\tilde{t}_i \ra b\chi^+_j$ can be 
written as $[q\equiv t$ or $b$, and we drop the indices of the 
neutralino/chargino states]
\beq
\Gamma_0( \tilde{t}_i \ra q \chi) = \frac{\alpha}{4\,m_{\tilde{t}_i}^3}
             \bigg[ ( {c_L^i}^2 + {c_R^i}^2 ) \, 
                     ( m_{\tilde{t}_i}^2 - m_{q}^2 - m_{\chi}^2 )
                    - 4 \, c_L^i\,c_R^i \, m_{q}\,m_{\chi}\, 
\epsilon_\chi
             \bigg] \,\lambda^{1/2}(m_{\tilde{t}_i}^2,m_{q}^2,m_{\chi}^2)   
\eeq
where $\lambda (x,y,z)=x^2+y^2+z^2-2\,(xy+xz+yz)$ is the usual two--body
phase space function and $\epsilon_\chi$ is the sign of the eigenvalue
of the neutralino $\chi$. The couplings $c_{L,R}^i$ for the neutral
current process, $\tilde{t}_i \ra t \chi^0$, are given by 
\begin{eqnarray}
\left\{ \begin{array}{c} c_R^1 \\  c_R^2 \end{array} \right\}
        &=&  b\,m_t\, \left\{ \begin{array}{c} \st{t} \\ \ct{t} \end{array} 
\right\}
         + f_L\, \left\{ \begin{array}{c} \ct{t} \\ -\st{t} \end{array} 
\right\} \nonumber \\
\left\{ \begin{array}{c} c_L^1 \\  c_L^2 \end{array} \right\}
        &=&  b\,m_t\, \left\{ \begin{array}{c} \ct{t} \\ -\st{t} \end{array} 
\right\}
         + f_R\, \left\{ \begin{array}{c} \st{t} \\ \ct{t} \end{array} \right\}
\end{eqnarray}
\beq
b & = & \frac{1}{\sqrt{2}\, M_W \sinb\,s_W} \; N_{j4} \nonumber \\ 
f_L & = & \sqrt{2}\left[ \frac{2}{3} \; N_{j1}'
              + \left(\frac{1}{2} - \frac{2}{3}\, s_W^2 \right)
                 \frac{1}{c_W s_W}\;N_{j2}' \right]  \nonumber \\ 
f_R & = &-\sqrt{2}\left[ \frac{2}{3} \; N_{j1}'
              - \frac{2}{3} \frac{s_W}{c_W} \; N_{j2}' \right] \ , 
\eeq
and for the charged current process, $\tilde{t}_i \ra b\chi^+$, 
\beq
\left\{ \begin{array}{c} c_L^1 \\ c_L^2 \end{array} \right\} & = &
   \frac{m_b\,U_{j2}}{\sqrt{2}\,s_W\,M_W\,\cosb}
\left\{ \begin{array}{c} -\ct{t} \\ \st{t} \end{array} \right\}
 \non \\
\left\{ \begin{array}{c} c_R^1 \\ c_R^2 \end{array} \right\} & = &
   \frac{V_{j1}}{s_W} \,
\left\{ \begin{array}{c} \ct{t} \\ -\st{t} \end{array} \right\}
 - \frac{m_t\,V_{j2}}{\sqrt{2}\,s_W\,M_W\,\sinb}
\left\{ \begin{array}{c} \st{t} \\ \ct{t} \end{array} \right\} \ . 
\eeq
In these equations, $\theta_t$ is the $\tilde{t}$ mixing angle [which as 
discussed previously can be expressed in terms of the Higgs--higgsino SUSY mass 
parameter $\mu$, $\tb$ and the soft--SUSY breaking trilinear coupling 
$A_t$] with $s_{\theta}=\sin\theta$, $c_{\theta}=\cos\theta$ etc.; 
$s_W^2=1-c_W^2\equiv \sin^2\theta_W$ and $N, U/V$ are the diagonalizing 
matrices for the neutralino and chargino states \cite{R10} with 
\beq
N'_{j1}= c_W N_{j1} +s_W N_{j2} \ \ \ , \ \ \
N'_{j2}= -s_W N_{j1} +c_W N_{j2} \ . 
\eeq
A similar expression eq.~(3) can be obtained for the neutral and charged 
decays of bottom squarks, $\tilde{b}_i \ra b \chi_j^0$ and $ \tilde{b} 
\ra t \chi_j^-$ 
\beq
\Gamma_0( \tilde{b}_i \ra q \chi) = \frac{\alpha}{4\,m_{\tilde{b}_i}^3}
             \bigg[ ( {c_L^i}^2 + {c_R^i}^2 ) \, 
                     ( m_{\tilde{b}_i}^2 - m_{q}^2 - m_{\chi}^2 )
                    - 4 \, c_L^i\,c_R^i \, m_{q}\,m_{\chi}\, 
\epsilon_\chi
             \bigg] \,\lambda^{1/2}(m_{\tilde{b}_i}^2,m_{q}^2,m_{\chi}^2)   
\eeq
with the couplings $c_{L,R}^i$ in the neutral decay $\tilde{b} \ra b 
\chi^0$ given by [$\theta_b$ is the $\tilde{b}$ mixing angle]
\begin{eqnarray}
\left\{ \begin{array}{c} c_R^1 \\  c_R^2 \end{array} \right\}
        &=&  b\,m_b\, \left\{ \begin{array}{c} \st{b} \\ \ct{b} \end{array} 
\right\}
         + f_L\, \left\{ \begin{array}{c} \ct{b} \\ -\st{b} \end{array} 
\right\} \nonumber \\
\left\{ \begin{array}{c} c_L^1 \\  c_L^2 \end{array} \right\}
        &=&  b\,m_b\, \left\{ \begin{array}{c} \ct{b} \\ -\st{b} \end{array} 
\right\}
         + f_R\, \left\{ \begin{array}{c} \st{b} \\ \ct{b} \end{array} \right\}
\end{eqnarray}
\beq
b & = & \frac{1}{\sqrt{2}\, M_W \cosb\,s_W} \; N_{j3} \nonumber \\ 
f_L & = & \sqrt{2}\left[ -\frac{1}{3} \; N_{j1}'
              + \left(-\frac{1}{2} +\frac{1}{3}\, s_W^2 \right)
                 \frac{1}{c_W s_W}\;N_{j2}' \right]  \nonumber \\ 
f_R & = &-\sqrt{2}\left[ -\frac{1}{3} \; N_{j1}'
              + \frac{1}{3} \frac{s_W}{c_W} \; N_{j2}' \right] \ , 
\eeq
and for the charged current process, $\tilde{b}_i \ra t\chi^-$, 
\beq
\left\{ \begin{array}{c} c_L^1 \\ c_L^2 \end{array} \right\} & = &
   \frac{m_t\,V_{j2}}{\sqrt{2}\,s_W\,M_W\,\sinb}
\left\{ \begin{array}{c} -\ct{b} \\ \st{b} \end{array} \right\}
 \non \\
\left\{ \begin{array}{c} c_R^1 \\ c_R^2 \end{array} \right\} & = &
   \frac{U_{j1}}{s_W} \,
\left\{ \begin{array}{c} \ct{b} \\ -\st{b} \end{array} \right\}
 - \frac{m_b\,U_{j2}}{\sqrt{2}\,s_W\,M_W\,\cosb}
\left\{ \begin{array}{c} \st{b} \\ \ct{b} \end{array} \right\} \ . 
\eeq
In the case where the mass of the final quark and the squark mixing
angle are neglected [as it is the case for the first and second
generation squarks], the decay widths simplify to 
\beq
\Gamma_0(\tilde{q}_i \ra q \chi) = \frac{\alpha}{4}\,m_{\tilde{q}_i}\,
\left( 1- \frac{m_{\chi}^2}{m_{\tilde{q}_i}^2}  \right)^2 f_i^2 
\eeq
where the $f_i$'s 
[with now $i=L,R$ since there is no squark mixing] in the case 
of the neutral decays, $\tilde{q} \ra q \chi^0$, are given in terms of
the quark isospin $I_{3L}^q$ and charge $e_q$, by
\beq
f_L & = & \sqrt{2}\left[ e_q \; N_{j1}'
              + \left(I_{3L}^q - e_q s_W^2 \right)
                 \frac{1}{c_W s_W}\;N_{j2}' \right]  \nonumber \\ 
f_R & = &-\sqrt{2}\left[ e_q \; N_{j1}'
              -e_q\, \frac{s_W}{c_W} \; N_{j2}' \right] \ , 
\eeq
while for the charged decays, $\tilde{q}  \ra q' \chi^+$ one has
for up--type (down--type) squarks:
\beq
f_L= V_{j1}/s_W \ (U_{j1}/s_W) \ \ , \ \ f_R=0 \ . 
\eeq

\subsection*{3. QCD corrections to Top Squark Decays}

The QCD corrections to the top squark decay width, eq.~(3), consist of virtual
corrections Figs.1a--d, and real corrections with an additional gluon
emitted off the initial $\tilde{t}$ or final $t$ [for the neutral decay] or
$b$ [for the charged decay] quark states, Fig.~1e. The ${\cal
O}(\alpha_s)$ virtual contributions can be split into gluon and gluino
exchange in the $q$--$\tilde{t}$--$\chi$ [$q=t,b$] vertex as well as
mixing diagrams and the $\tilde{t}$ and $t/b$ wave function
renormalization constants. The renormalization of the $q$--$\tilde{t}$--$\chi$
coupling is achieved by renormalizing the top/bottom quark masses and
the $\tilde{t}$ mixing angle. We will use the dimensional reduction
scheme\footnote{The quark mass and wave-function counterterms will
be different in the dimensional regularization \cite{R11a} and dimensional 
reduction schemes \cite{R11}. Since dimensional reduction is
the scheme which preserves supersymmetry, we will present our results in
this scheme.} to regularize the ultraviolet divergencies, and 
a fictitious gluon mass $\lambda$ is introduced to regularize the infrared
divergencies. 

\subsubsection*{3.1 Virtual Corrections}

The QCD virtual corrections to the $\tilde{t}_i$--$\chi$--$q$ interaction 
vertex can be cast into the form 
\beq 
\delta \Gamma^i = ie \ \frac{\alpha_s}{3\pi} \, \sum_{j=g, 
\tilde{g}, {\rm mix}, {\rm ct} } \left[ G_{j,L}^i P_L + G_{j,R}^i P_R \right]
\eeq 
where $G^i_{g}, \,G^i_{\tilde{g}}, \,G^i_{\rm mix}$ and $G^i_{\rm ct}$ denote 
the gluon and gluino exchanges in the vertex, and the mixing and counterterm
contributions, respectively. \s

The contribution of the gluonic exchange [Fig.~1a] can be written as
\beq
G^i_{g,L,R} = c_{L,R}^i \, F_1^i + c_{R,L}^i \,F_2^i 
\eeq
with the form factors $F^i_{1,2}$ given by 
\beq
F_1^i & = &  B_0 
 + 2 \, m_{q}^2 \, C_0 - 2 \, m_{\tilde{t}_i}^2 \, (C_{11}-C_{12}) 
 + 2 \, m_{\chi}^2 \, C_{11} \nonumber \\
F_2^i & = & -2 \, m_{q} \, m_{\chi} \, (C_0+C_{11})
\eeq
with $q \equiv t$ for the neutral and $q \equiv b$ for the charged 
decays; the two and three--point Passarino--Veltman functions, $B_0
\equiv B_0(m_{\tilde{t}_i}^2,\lambda,m_{\tilde{t}_i})$ and
$C_{..} \equiv C_{..}(m_{q}^2, m_{\tilde{t}_i}^2, m_{\chi}^2,$ $m_{q}^2,
\lambda^2,m_{\tilde{t}_i}^2)$ can be found in Ref.~\cite{R12}. \s 

The gluino exchange contributions [Fig.~1b], are given by 
\beq
G_{\tilde{g},L,R}^i & = & -2 \sum_{k=1,2} \, 
  d_{L,R}^k \bigg[
        (v_{\tilde{q}}^k v_{\tilde{t}}^i+a_{\tilde{q}}^k a_{\tilde{t}}^i) 
F_4^{ik}
    \mp (a_{\tilde{q}}^k v_{\tilde{t}}^i+v_{\tilde{q}}^k a_{\tilde{t}}^i) 
F_5^{ik} 
\non \\
& & \hspace{1.5cm}
      + (v_{\tilde{q}}^k v_{\tilde{t}}^i-a_{\tilde{q}}^k a_{\tilde{t}}^i) 
F_6^{ik}
    \mp (a_{\tilde{q}}^k v_{\tilde{t}}^i-v_{\tilde{q}}^k a_{\tilde{t}}^i) 
F_7^{ik}
        \bigg]
\non \\
& & \hspace{1.5cm}
+ d_{R,L}^k \bigg[
        (v_{\tilde{q}}^k v_{\tilde{t}}^i+a_{\tilde{q}}^k a_{\tilde{t}}^i) 
F_1^{ik}
    \mp (a_{\tilde{q}}^k v_{\tilde{t}}^i+v_{\tilde{q}}^k a_{\tilde{t}}^i) 
F_1^{ik}
\non \\
& & \hspace{1.5cm}
      + (v_{\tilde{q}}^k v_{\tilde{t}}^i-a_{\tilde{q}}^k a_{\tilde{t}}^i) 
F_2^{ik}
    \mp (a_{\tilde{q}}^k v_{\tilde{t}}^i-v_{\tilde{q}}^k a_{\tilde{t}}^i) 
F_3^{ik}
        \bigg]
\eeq
with again $q=t$ for the neutral decay and $q=b$ for the charged one; the 
form factors $F^{ik}_{1,..,7}$ read 
\beq
    F_1^{ik}     & = & m_{\tilde{g}}\,m_{\chi}\, [C_0+C_{12}] \non \\
    F_{2,3}^{ik} & = & m_{\chi}\,
                    [\pm  m_{q}\, (C_0+C_{11})+m_t\,C_{12}] \non \\
    F_{4,5}^{ik} & = & m_{\tilde{g}}\, [m_t \,C_0 
                          \pm m_{q}\,(C_{11}-C_{12})] \non \\
    F_{6,7}^{ik} & = & m_{\tilde{q}_k}^2\,C_0 \pm 
                    m_t \,m_{q}\, [C_0+C_{11}-C_{12}]
                  +m_{q}^2\, [C_{11}-C_{12}] +m_{\chi}^2\,C_{12}+B_0
\eeq
with the two-- and three--point functions 
$B_0\equiv B_0(m_{\tilde{t}_i}^2,m_{\tilde{g}},m_t)$ and $C_{..} \equiv 
C_{..} (m_{q}^2,m_{\tilde{t}_i}^2,m_{\chi}^2,m_{\tilde{q}}^2,$ 
$m_{\tilde{g}}^2, m_t^2)$. The couplings $d_{R,L}^k$ are given by
\beq
d_{L,R}^k \hspace{0.3cm}  = c_{R,L}^k  \ 
\eeq
for neutralinos, while for the charginos one has
\beq 
\left\{\begin{array}{c} d_L^1 \\ d_L^2 \end{array} \right\}
 & = &
   \frac{U_{j1}}{s_W}\,
\left\{\begin{array}{c} \ct{b} \\ -\st{b} \end{array} \right\}
  -\frac{m_b\,U_{j2}}{\sqrt{2}\,s_W\,M_W\,\cosb}\,
\left\{\begin{array}{c} \st{b} \\ \ct{b} \end{array} \right\} \non \\
\left\{\begin{array}{c} d_R^1 \\ d_R^2 \end{array} \right\}
 & = &
   \frac{m_t\,V_{j2}}{\sqrt{2}\,s_W\,M_W\,\sinb}\,
\left\{\begin{array}{c} -\ct{b} \\ \st{b} \end{array} \right\} \ . 
\eeq
The $v_{\tilde{q}}^i$ and $a_{\tilde{q}}^i$ couplings read
\beq
      v_{\tilde{q}}^1 & = & {\textstyle\frac{1}{2}}\,( \ct{q}-\st{q} ) \ , 
           \hspace{1.cm}
      v_{\tilde{q}}^2 \; = \; {\textstyle - \frac{1}{2}}\,( \ct{q}+\st{q} ) 
\ , \non \\
      a_{\tilde{q}}^1 & = & {\textstyle\frac{1}{2}}\,( \ct{q}+\st{q} )
           \ , \hspace{1.cm}
      a_{\tilde{q}}^2 \; = \; {\textstyle\frac{1}{2}}\,( \ct{q}-\st{q} ) \ .
\eeq

\vspace*{3mm} 

Finally, the mixing contributions due to the diagrams Fig.~1c, yield 
the expressions 
\beq
G_{\rm mix,L,R}^i & = & \frac{(-1)^i\,(\delta_{1i}\,c_{L,R}^2 
                                 + \delta_{2i}\,c_{L,R}^1)}
                       {m_{\tilde{t}_1}^2-m_{\tilde{t}_2}^2} \,
\bigg[ 4 m_t \,m_{\tilde{g}}\, c_{2 \theta_t}\,B_0(m_{\tilde{t}_i}^2,
      m_t,m_{\tilde{g}}) \non \\   
& & \hspace{4.5cm}   + \, c_{2 \theta_t} s_{2\theta_t} (
        A_0(m_{\tilde{t}_2}^2)-  A_0(m_{\tilde{t}_1}^2) ) \bigg] \ . 
\eeq
Therein, $A_0$ is the Passarino--Veltman one--point function. 
Note that all these contributions are the same in both the dimensional
reduction and dimensional regularization schemes. 

\subsubsection*{3.2 Counterterms}

The counterterm contributions in eq.~(15) are due to the $\tilde{t}$ 
and $t/b$ 
wave function renormalizations [Fig.~1d] as well as the renormalization of the 
quark mass $m_t$ or $m_b$ and the mixing angle $\theta_t$, which appear in the 
Born couplings. \s

For the neutral decay process, $\tilde{t}_i \rightarrow t\chi^0_j$, the
counterterm contribution is given by 
\beq
G^{1,2}_{\rm ct,L} & = & \frac{1}{2}\,c^{1,2}_L\,( \delta Z^t_R
                                             + \delta Z_{\tilde{t}_{1,2}}) 
                    + b \, \{\ct{t},-\st{t}\} \, \delta m_t
                    - b\,m_t \, \{\st{t},\ct{t}\} \, \delta \theta_t 
                    + f_R \, \{\ct{t},-\st{t}\} \, \delta \theta_t  \non \\
G^{1,2}_{\rm ct,R} & = & \frac{1}{2}\,c^{1,2}_R\,( \delta Z^t_L
                                             + \delta Z_{\tilde{t}_{1,2}}) 
                    + b \, \{\st{t},\ct{t}\} \, \delta m_t
                    + b\,m_t \, \{\ct{t},-\st{t}\} \, \delta \theta_t
                    - f_L \, \{\st{t},\ct{t}\} \, \delta \theta_t
\ , \non \\ && \eeq
whereas for the charged current process, $\tilde{t}_i \rightarrow 
b\chi^+_j$, one obtains, 
\beq
G^{1,2}_{\rm ct,L} & = & \frac{1}{2}\,c^{1,2}_L\, \left[ \delta Z^b_R 
                        + \delta Z_{\tilde{t}_{1,2}} 
                        + 2\,\frac{\delta m_b}{m_b} \right] 
                        + \frac{ m_b U_{j2}}{\sqrt{2}\,s_W\,M_W\,\cos\beta}
                              \,\{\st{t},\ct{t}\}\,\delta \theta_t  \non \\
G^{1,2}_{\rm ct,R} & = & \frac{1}{2}\,c^{1,2}_R\, \left[ \delta Z^b_L 
                       + \delta Z_{\tilde{t}_{1,2}} \right] 
                       - \,\frac{ \delta m_t \, V_{j2}}{\sqrt{2}\,s_W\,M_W\,
                        \sin\beta}
                          \,\{\st{t},\ct{t}\} \non \\  &  & \vspace{0.5cm}
                       - \frac{V_{j1}}{s_W}\,\{\st{t},\ct{t}\}\,\delta \theta_t
                       - \frac{m_t V_{j2}}{\sqrt{2}\,s_W\,M_W\,\sin\beta}
                          \,\{\ct{t},-\st{t}\}\,\delta \theta_t \ . 
\eeq
In the on--shell scheme, the quark and squark masses are defined as 
the poles of the propagators and the wave--function renormalization
constants  follow from the residues at the poles; the corresponding 
counterterms are given by  (see also Refs.~\cite{R6,R8})
\beq 
\frac{\delta m_q}{m_q} & = & \frac{1}{2}
\bigg[ \Sigma^q_R(m_q^2)+\Sigma^q_L(m_q^2)\bigg] 
+ \Sigma^q_S(m_q^2)   \nonumber \\
\delta Z^q_L & = & - \Sigma^q_L(m_q^2)
                   - m_q^2 \bigg[ {\Sigma^q_L}^{\prime}(m_q^2)
                   +{\Sigma^q_R}^{\prime}(m_q^2)+2\,{\Sigma^q_S}^{\prime}(m_q^2)
                           \bigg]   \nonumber \\
\delta Z^q_R & = & - \Sigma^q_R(m_q^2) 
                   - m_q^2 \bigg[ {\Sigma^q_L}^{\prime}(m_q^2)
                   +{\Sigma^q_R}^{\prime}(m_q^2)+2\,{\Sigma^q_S}^{\prime}(m_q^2)
                           \bigg]   \nonumber \\
\delta Z_{\tilde{t}_i} & = & - \left(\Sigma_{\tilde{t}}^{ii}
\right)'(m_{\tilde{t}_i}^2)
\eeq
In the dimensional reduction scheme, the self--energies $\Sigma$ and their 
derivatives $\Sigma'$, up to a factor $\alpha_s /3\pi$ which has been 
factorized out, are given by \cite{R6,R8}
\beq
\Sigma^q_L(k^2) & = &  - \bigg[ 2 \,B_1(k^2,m_q,\lambda)
            + (1+c_{2 \theta_q}) B_1(k^2,m_{\tilde{g}},m_{\tilde{q}_1})
            + (1-c_{2 \theta_q}) B_1(k^2,m_{\tilde{g}},m_{\tilde{q}_2}) \bigg]
   \nonumber \\
\Sigma^q_R(k^2) & = &  - \bigg[ 2 \,B_1(k^2,m_q,\lambda)
            + (1-c_{2 \theta_q}) B_1(k^2,m_{\tilde{g}},m_{\tilde{q}_1})
            + (1+c_{2 \theta_q}) B_1(k^2,m_{\tilde{g}},m_{\tilde{q}_2}) \bigg]
   \nonumber \\
\Sigma^q_S(k^2) & = &  -    \bigg[ 4 \,B_0(k^2,m_q,\lambda)
      + \frac{m_{\tilde{g}}}{m_q}\,s_{2 \theta_q}\,
         ( B_0(k^2,m_{\tilde{g}},m_{\tilde{q}_1})
         - B_0(k^2,m_{\tilde{g}},m_{\tilde{q}_2}) ) \bigg]
   \nonumber \\
(\Sigma_{\tilde{t}}^{ii})'(k^2) & = & - 2 \bigg[
- 2\,B_1(k^2,m_{\tilde{t}_i},\lambda) 
- 2\,k^2 \,B_1'(k^2,m_{\tilde{t}_i},\lambda) 
+ (m_t^2+m_{\tilde{g}}^2-k^2)\,B_0'(k^2,m_t,m_{\tilde{g}}) 
\non \\ & & \hspace{0.8cm}
- \,B_0(k^2,m_t,m_{\tilde{g}}) + (-1)^i\,2\,s_{2 \theta}\,m_t\,m_{\tilde{g}} 
B_0'(k^2,m_t,m_{\tilde{g}}) \bigg] \ . 
\eeq
Using dimensional regularization, the quark self--energies differ from the 
previous expressions by a constant; in terms of the their values in the
dimensional reduction scheme, they are given by
\beq
\left. \Sigma_{L,R}^q \right|_{\rm dim.~reg.} = \Sigma_{L,R}^q -2
\ \ , \ \
\left. \Sigma_{S}^q \right|_{\rm dim.~reg.} = \Sigma_{S}^q + 2 \ . 
\eeq

Finally, we need a prescription to renormalize the $\tilde{t}$ mixing angle
$\theta_t$. Following Ref.~\cite{R13}, we choose this condition in such a way 
that it cancels exactly the mixing contributions eq.~(23) for the decay
$\tilde{t_2} \ra t \chi^0$
\beq
\delta\theta_t & = & \frac{1} {m_{\tilde{t}_1}^2-m_{\tilde{t}_2}^2} 
\left[4 \, m_t \,m_{\tilde{g}} \,c_{2 \theta_t} \,B_0(m_{\tilde{t}_2}^2,m_t,
m_{\tilde{g}}) + c_{2 \theta_t} s_{2\theta_t} (A_0(m_{\tilde{t}_2}^2)- 
A_0(m_{\tilde{t}_1}^2) ) \right] \ . 
\eeq

Alternatively, since the lightest top squark $\tilde{t}_1$ can be lighter 
than the top quark and then is more likely to be discovered first 
in the top decays $t \ra \tilde{t}_1 \chi_0$, one can choose the 
renormalization condition such that 
the mixing contributions are cancelled in the latter process; this
leads to a counterterm similar to eq.~(29) but with $B_0(m_{\tilde{t}_2}^2,m_t,
m_{\tilde{g}})$ replaced by $B_0(m_{\tilde{t}_1}^2,m_t, m_{\tilde{g}})$.
The difference between the two renormalization conditions,
\beq
\Delta \delta\theta_t = \frac{4 m_t \,m_{\tilde{g}} \,c_{2 \theta_t}}
{m_{\tilde{t}_1}^2-m_{\tilde{t}_2}^2} 
\left[ B_0(m_{\tilde{t}_1}^2,m_t,m_{\tilde{g}}) 
-      B_0(m_{\tilde{t}_2}^2,m_t,m_{\tilde{g}}) \right]
\eeq
is, however, very small numerically. Indeed, if $m_{\tilde{t}_1}$ is a few GeV 
away from $m_{\tilde{t}_2}$, one has $\theta_t \simeq -\pi/4$ and therefore 
$c_{2 \theta_t} \sim 0$, leading to a difference which is less than one
permille for the scenario of Figs.~2a/b. For degenerate top squarks, one 
has $\Delta \delta \theta =4m_t m_{\tilde{g}} c_{2 \theta_t} B_0'
(m_{\tilde{t}_2}^2,m_t,m_{\tilde{g}})$ which is also very small numerically
[less than $\sim 1\% $ for the scenarios of Fig.~2.] \s

The complete virtual corrections to the $\tilde{t}_i \ra q \chi$ 
decay width is then given by
\begin{eqnarray}
\Gamma^V(\tilde{t}_i \rightarrow q \chi) & = &
\frac{\alpha}{6 \, m_{\tilde{t}_i}^3} \frac{\alpha_s}{\pi} 
       \;   \mbox{Re} \; \bigg\{
       (c_L^i \, G_L^i + c_R^i \, G_R^i)\,
       ( m_{\tilde{t}_i}^2 - m_{q}^2 - m_{\chi}^2 ) 
                    \nonumber \\ 
 & & \hspace{1.6cm} 
     - \; 2 \; ( c_L^i \, G_R^i + c_R^i \, G_L^i ) 
              \, m_{q} \, m_{\chi} \epsilon_\chi \, 
                    \bigg\} \,
    \lambda^{1/2}(m_{\tilde{t}_i}^2,m_{q}^2,m_{\chi}^2) \ .
\end{eqnarray}

The sum of all virtual contributions including the counterterms are
ultraviolet finite as it should be, but they are still infrared divergent; 
the infrared divergencies will be cancelled after adding the real corrections. 

\subsubsection*{3.3 Real Corrections}

The contributions to the squark decay widths from the real corrections, 
with an additional gluon emitted from the initial $\tilde{t}$ or final 
$t/b$ states, can be cast into the form 
\beq
\Gamma_{\rm real}^i & = & \frac{2\,\alpha}{3 \, m_{\tilde{t}_i}} 
\frac{\alpha_s}{\pi}
\bigg\{ 
  8 \; c_L^i \, c_R^i \; m_{q} \, m_{\chi} \epsilon_\chi \,  
 \big[\; ( m_{\tilde{t}_i}^2 + m_{q}^2 - m_{\chi}^2) \, I_{01}
       +  m_{\tilde{t}_i}^2 \, I_{00}
       +  m_{q}^2 \, I_{11}
       +  I_0  +  I_1  \big]
   \nonumber \\
 & & \hspace{1.6cm} +\; ({c_L^i}^2+{c_R^i}^2) \,
  \big[\;  2 \, ( m_{q}^2 + m_{\chi}^2 - m_{\tilde{t}_i}^2 )
                      \, ( m_{\tilde{t}_i}^2 \, I_{00} + m_{q}^2 \, I_{11} 
                          + I_0 + I_1 )
   \nonumber \\ 
 & & \hspace{4.1cm}
       + 2 \, ( m_{q}^4 - \; ( m_{\chi}^2 - m_{\tilde{t}_i}^2 )^2 ) \, I_{01}
       - I
       - I_1^0 \big]
\bigg\} 
\eeq
where the phase space integrals $ I(m_{\tilde{t}_i},m_{q},m_{\chi})
\equiv I $ are given by \cite{R14}
\beq
I_{00} & = & \frac{1}{4\,m_{\tilde{t}_i}^4}\bigg[ \kappa \, \ln \bigg(
    \frac{\kappa^2}{\lambda\,m_{\tilde{t}_i}\,m_{q}\,m_{\chi}}\bigg)
    -\kappa-(m_{q}^2-m_{\chi}^2) \ln
     \bigg(\frac{\beta_1}{\beta_2}\bigg)-m_{\tilde{t}_i}^2\,\ln (\beta_0) 
\bigg] \non \\ 
I_{11} & = & \frac{1}{4\,m_{q}^2\,m_{\tilde{t}_i}^2}\bigg[ \kappa \, \ln 
\bigg(
    \frac{\kappa^2}{\lambda\,m_{\tilde{t}_i}\,m_{q}\,m_{\chi}}\bigg)
    -\kappa-(m_{\tilde{t}_i}^2-m_{\chi}^2)\ln 
  \bigg(\frac{\beta_0}{\beta_2}\bigg)-m_{q}^2\,\ln (\beta_1)
    \bigg] \non \\
I_{01} & = & \frac{1}{4\,m_{\tilde{t}_i}^2}
              \bigg[ -2\,\ln \bigg(\frac{\lambda\,m_{\tilde{t}_i}\,
              m_{q}\,m_{\chi}}{\kappa^2} \bigg)\,\ln (\beta_2) 
          + 2\,\ln^2(\beta_2) - \ln^2(\beta_0) - \ln^2(\beta_1) \non \\
    &  & + 2\,\mbox{Li}_2\,(1-\beta_2^2) - \mbox{Li}_2 \,(1-\beta_0^2)
         - \mbox{Li}_2\,(1-\beta_1^2) \bigg] \non \\
I & = & \frac{1}{4\,m_{\tilde{t}_i}^2} \bigg[ \frac{\kappa}{2}(m_{\tilde{t}_i}^2
                     +m_{q}^2+m_{\chi}^2)
       +2\,m_{\tilde{t}_i}^2\,m_{q}^2\,\ln (\beta_2)
       +2\,m_{\tilde{t}_i}^2\,m_{\chi}^2\,\ln (\beta_1)
       +2\,m_{q}^2\,m_{\chi}^2\,\ln (\beta_0) \bigg] \non \\
I_0 & = & \frac{1}{4\,m_{\tilde{t}_i}^2} \bigg[ -2\,m_{q}^2\,\ln (\beta_2)
         -2\,m_{\chi}^2\,\ln (\beta_1)-\kappa \bigg] \non \\
I_1 & = & \frac{1}{4\,m_{\tilde{t}_i}^2}\bigg[ -2\,m_{\tilde{t}_i}^2\,
\ln (\beta_2)
         -2\,m_{\chi}^2\,\ln (\beta_0)-\kappa \bigg] \non \\
I_1^0 & = & \frac{1}{4\,m_{\tilde{t}_i}^2}\bigg[ m_{\tilde{t}_i}^4 \,
\ln (\beta_2)
          -m_{\chi}^2 \,(2\,m_{q}^2-2\,m_{\tilde{t}_i}^2
          +m_{\chi}^2) \, \ln (\beta_0)
          -\frac{\kappa}{4}\,(m_{q}^2-3\,m_{\tilde{t}_i}^2+5\,m_{\chi}^2) 
\bigg] \ . 
\eeq
with $\kappa = \lambda^{1/2}(m_{\tilde{t}_i}^2,m_{q},m_{\chi})$ and
\begin{equation} 
\beta_0 = \frac{m_{\tilde{t}_i}^2-m_{q}^2-m_{\chi}^2+\kappa}
    {2\,m_{q}\,m_{\chi}},\;\;
\beta_1 = \frac{m_{\tilde{t}_i}^2-m_{q}^2+m_{\chi}^2-\kappa}
    {2\,m_{\tilde{t}_i}\,m_{\chi}},\;\;
\beta_2 = \frac{m_{\tilde{t}_i}^2+m_{q}^2-m_{\chi}^2-\kappa}
    {2\,m_{\tilde{t}_i}\,m_{q}} \ .
\end{equation}

\bigskip

\nn Our analytical results agree with the results obtained recently in
Ref.~\cite{R8}.

\subsection*{4. QCD corrections to other squark decays}

\subsubsection*{4.1 Bottom Squark Decays}

In the case of the bottom squark decays, $\tilde{b}_i \ra b \chi^0$ and
$\tilde{b}_i \ra t \chi^-$, the analytical expressions of the QCD
corrections are just the same as in the previous section once the proper
changes of the squark [$m_{\tilde{t}_i} \ra m_{\tilde{b}_i}$], the quark
$[q\equiv b$ and $q\equiv t$ for the neutral and charged decays] masses
and the mixing angles $[\theta_t \ra \theta_b$] are performed. The couplings 
for $\tilde{b}$ decays are as given in section 2: for the $d^k_{L,R}$
couplings, one has in the case of the neutral decay 
$\tilde{b}_i \ra b \chi^0$ 
\beq
d_{L,R}^k \hspace{0.3cm}  = c_{R,L}^k \ , 
\eeq
with $c_{L,R}^k$ of eq.~(11), while in the charged decay $\tilde{b}_i 
\ra t \chi^-$, they read
\beq 
\left\{\begin{array}{c} d_L^1 \\ d_L^2 \end{array} \right\}
 & = &
   \frac{V_{j1}}{s_W}\,
\left\{\begin{array}{c} \ct{t} \\ -\st{t} \end{array} \right\}
  -\frac{m_t\,V_{j2}}{\sqrt{2}\,s_W\,M_W\,\sinb}\,
\left\{\begin{array}{c} \st{t} \\ \ct{t} \end{array} \right\} \non \\
\left\{\begin{array}{c} d_R^1 \\ d_R^2 \end{array} \right\}
 & = &
   \frac{m_b\,U_{j2}}{\sqrt{2}\,s_W\,M_W\,\cosb}\,
\left\{\begin{array}{c} -\ct{t} \\ \st{t} \end{array} \right\} \ .
\eeq
The counterterm contributions are the same as in eq.~(24) with the 
change $(t, \tilde{t}) \ra (b, \tilde{b})$ in the neutral
decay; in the charged decay mode they are different due to different
couplings (see also Refs.~\cite{R6,R8}):
\beq
G^{1,2}_{\rm ct,L} & = & \frac{1}{2}\,c^{1,2}_L\, \left[ \delta Z^t_R 
                        + \delta Z_{\tilde{b}_{1,2}} 
                        + 2\,\frac{\delta m_t}{m_t} \right] 
                        + \frac{ m_t V_{j2}}{\sqrt{2}\,s_W\,M_W\,\sin\beta}
                              \,\{\st{b},\ct{b}\}\,\delta \theta_b  \non \\
G^{1,2}_{\rm ct,R} & = & \frac{1}{2}\,c^{1,2}_R\, \left[ \delta Z^t_L 
                       + \delta Z_{\tilde{b}_{1,2}} \right] 
                       - \,\frac{ \delta m_b \, U_{j2}}{\sqrt{2}\,s_W\,M_W\,
                        \cos\beta}
                          \,\{\st{b},\ct{b}\} \non \\  &  & \vspace{0.5cm}
                       - \frac{U_{j1}}{s_W}\,\{\st{b},\ct{b}\}\,\delta \theta_b
                       - \frac{m_b U_{j2}}{\sqrt{2}\,s_W\,M_W\,\cos\beta}
                          \,\{\ct{b},-\st{b}\}\,\delta \theta_b \ .
\eeq
where again the $c_{L,R}^k$ are given by eq.~(11). Except for very large 
values of $\tb$, the $\tilde{b}$ mixing angle [as well as the bottom quark 
mass] can be set to zero and the analytical expressions simplify 
considerably\footnote{In the absence of mixing, the left-- and 
right--handed bottom squarks are, to a very good approximation, 
degenerate if $M_{\tilde{q}_L} = M_{\tilde{q}_R}$. In the rest of the 
discussion, ${\tilde{b}_L}$ and ${\tilde{b}_R}$ [and a {\it fortiori} 
the partners of the light quarks ${\tilde{q}_L}$ and ${\tilde{q}_R}$] 
will be considered as degenerate.}. 
The case of the neutral decay $\tilde{b} \ra b \chi^0$ is 
even simpler since one can also neglect the mass of the final $b$ quark. 
In fact, the latter situation corresponds to the case of decays of first 
and second generation squarks into light quarks and charginos/neutralinos, 
which will be discussed now. 

\subsubsection*{4.2 Light Quark Partners Decays}

Neglecting the squark mixing angle as well as the mass of the final
quarks, the virtual corrections of the processes $\tilde{q}_i \ra q
\chi$ [where the subscript $i$ stands now for the chirality of the
squark, since in the absence of squark mixing one has $\tilde{q}_{L,R}
=\tilde{q}_{1,2}$] are given by the sum  of the gluon and gluino 
exchange vertices and the wave--function counterterm, plus the real 
correction. The total width can then be written as
\beq 
\Gamma^i = \Gamma^i_0
\bigg[ 1 \,+ \,    \frac{4}{3} \frac{\alpha_s}{\pi} \, 
 \left( F_{\rm g}+ F_{\rm \tilde{g}}+ F_{\rm ct} + F_{\rm r} \right) \bigg]
\eeq 
where the decay width in the Born approximation $\Gamma^i_0$ has been given 
in eq.~(12).
In terms of the ratio $\kappa= m_{\chi}^2/m_{\tilde{q}}^2$, the gluon 
exchange corrections are given by [$\Delta =1/(4-n)$ with $n$ the 
space-time dimension, and $\mu$ is the renormalization scale]
\beq
F_{\rm g} &=& \frac{\Delta}{2} + 1 - \frac{1}{2} \ln \frac{ m_ {\tilde{q}}^2 }
{\mu^2} -\frac{1}{4} \ln^2 \frac{ \lambda^2/ m_ {\tilde{q}}^2 } {(1-\kappa)^2 }
- \ln \frac{ \lambda^2/ m_ {\tilde{q}}^2 } {1-\kappa}
-{\rm Li_{2}}(\kappa) \ .
\eeq
The gluino exchange contribution, with $\gamma= m_{\tilde{g}}^2
/m_{\tilde{q}}^2$, is given by
\beq
F_{\rm \tilde{g}} = \sqrt{ \kappa \gamma} \left[ \frac{1}{ \kappa} \ln 
(1-\kappa)+ \frac{1}{1-\kappa} \left[ \gamma \ln \gamma -(\gamma-1)
\ln (\gamma-1) \right] + \frac{ \kappa +\gamma -2}{(1-\kappa)^2} \, I \, 
\right] 
\eeq
with 
\beq
I  & \equiv & \frac{1}{m_{\tilde{q}_i}^2\,(1-\kappa)}\,
C_0 (0,m_{\tilde{q}}^2, m_{\chi}^2,m_{\tilde{q}}^2,m_{\tilde{g}}^2, 0) \ .
\eeq
In terms of dilogarithms, the function $I$ is given for 
$\kappa \gamma <1$ by 
\beq
I= {\rm Li_{2}} \left( \frac{\gamma-1}{\gamma \kappa-1} \right)
 - {\rm Li_{2}} \left( \kappa \frac{\gamma-1}{\gamma \kappa-1} \right)
 - {\rm Li_{2}} \left( \frac{\gamma+\kappa-2}{\gamma \kappa-1} \right)
 + {\rm Li_{2}} \left( \kappa \frac{\gamma+\kappa-2}{\gamma \kappa-1} \right) 
\eeq
and for $\kappa \gamma > 1$ one has
\beq
I & = &-{\rm Li_{2}}\left( \frac{\gamma \kappa-1}{\gamma-1}  \right)
       +{\rm Li_{2}}\left( \frac{\gamma \kappa-1}{\gamma+\kappa-2}  \right)
       +{\rm Li_{2}}\left( \frac{\gamma \kappa-1}{\kappa(\gamma-1)}  \right)
       -{\rm Li_{2}}\left( \frac{\gamma \kappa-1}{\kappa(\gamma+\kappa-2)}  
\right)          \non \\
  &   & -\ln (\kappa)\,\ln \frac{\gamma+\kappa-2}{\gamma-1} \ .
\eeq
The counterterm contribution, consisting of the sum of the squark and 
quark wave--function renormalization constants, reads
\beq
F_{\rm ct} &=& - \frac{\Delta}{2} + \frac{\gamma}{4\,(1-\gamma)} - 
\frac{\gamma}{2}
 - \frac{15}{8}
 + \frac{1}{2} \ln \frac{ m_ {\tilde{q}}^2}{\mu^2}
 - \frac{1}{4} \ln \frac{ \lambda^2}{ m_ {\tilde{q}}^2 }  \non \\ & & 
 - \frac{1}{2}(\gamma^2-1) \ln \frac{\gamma -1}{\gamma}
 + \frac{1}{4}\left[ \frac{2\,\gamma-1}{(1-\gamma)^2}+3  \right] \ln \gamma
\ . \eeq
Finally, the real corrections with massless quarks in the final 
state contribute 
\beq
F_{\rm r}  &=&
  \frac{1}{4} \ln^2 \frac{ \lambda^2/ m_{\tilde{q}}^2 }{(1-\kappa)^2}
+ \frac{5}{4} \ln   \frac{ \lambda^2/ m_{\tilde{q}}^2 }{(1-\kappa)^2}
- \frac{\kappa\,(4-3 \kappa)}{4\,(1-\kappa)^2}\ln \kappa
\non \\ & &
- {\rm Li_{2}}(\kappa) -\ln \kappa \ln (1-\kappa)  
- \frac{3 \kappa-5}{8\,(\kappa-1)} - \frac{\pi^2}{3} + 4 \ .
\eeq
We see explicitly that the ultraviolet divergences $\Delta/2$ and the
scale $\mu$ cancel
when $F^i_g$ and $F^i_{\rm ct}$ are added, and that the infrared 
divergences $\ln^2(\lambda^2/ m_{\tilde{q}}^2)$ and $\ln (\lambda^2/ 
m_{\tilde{q}}^2)$ disappear when $F_g$, $F_{\rm ct}$ and $F_{\rm 
r}$ are summed. The gluino exchange contribution eq.~(40) does not contain
any ultraviolet or infrared divergences. The total correction in 
eq.~(38) then reads
\beq
F_{\rm tot} & = &  F_{\rm g}+ F_{\rm \tilde{g}}+ F_{\rm ct} + F_{\rm r} \non \\
            & = & - \frac{1}{8}\left( \frac{4\,\gamma^2-27\,\gamma+25}{\gamma-1}
                                    + \frac{3\,\kappa-5}{\kappa-1} \right)
  - \frac{\pi^2}{3} - 2\, {\rm Li_2}(\kappa)
  - \frac{1}{2}\,(\gamma^2-1)\,\ln \frac{\gamma-1}{\gamma} \non \\
   &  & 
  + \frac{3\,\gamma^2-4\,\gamma+2}{4\,(1-\gamma)^2}\,\ln \gamma
  - \frac{3}{2}\,\ln (1- \kappa) 
  + \frac{3\,\kappa^2-4\,\kappa}{4\,(\kappa-1)^2}\,\ln \kappa 
  - \ln \kappa\,\ln (1-\kappa) \non \\ 
   &  & +\sqrt{ \kappa \gamma} \left[ \frac{1}{ \kappa} \ln 
(1-\kappa)+ \frac{1}{1-\kappa} \left[ \gamma \ln \gamma -(\gamma-1)
\ln (\gamma-1) \right] + \frac{ \kappa +\gamma -2}{(1-\kappa)^2} \, I \, 
\right] . 
\eeq

\smallskip

In the limit where the mass of the final neutralino or chargino is much
smaller than the mass of the initial squark, the analytical expression
of the QCD correction further simplifies:
\beq
F_{\rm tot}= \frac{3 \gamma^2-4\gamma+2}{4(\gamma-1)^2} \ln \gamma 
- \frac{1}{2}
(\gamma^2-1) \ln \frac{\gamma-1}{\gamma} -\frac{2 \gamma^2-11 \gamma
+10}{4(\gamma-1)} -\frac{\pi^2}{3} \ . 
\eeq
Note the explicit logarithmic dependence on the gluino mass in the 
correction. This logarithmic behaviour, leading
to a non-decoupling of the gluinos for very large masses, 
\beq
F_{\rm tot} = \frac{3}{4} \ln \frac{ m_{\tilde{g}}^2} {m_{\tilde{q}}^2}
+\frac{5}{2} -\frac{\pi^2}{3} \ \ \ {\rm for} \ \ 
m_{\tilde{g}} \gg m_{\tilde{q}}
\eeq
is due to the wave function renormalization and is a consequence of the
breakdown of SUSY as discussed in Ref.~\cite{R4}. Had we chosen
the $\overline{\rm MS}$ scheme when renormalizing the squark/quark wave
functions [i.e. subtracting only the poles and the related constants in
the expression eq.~(23)] we would have been left with contributions which 
increase linearly with the gluino mass. \s 

Our analytical results in the case of massless final quarks agree with the
corresponding results obtained in Refs.\cite{R4,R5}, where the QCD
corrections to the decay of a squark into a massless quark and a photino
have been derived, after correcting the sign of $F_{\tilde{g}}$ in 
Ref.~\cite{R4}; see also the discussion given in Ref.~\cite{R5}.

\subsection*{5. Numerical Analysis and Discussion}

In the numerical analysis of the QCD corrections to squark decays, we
will choose $m_t=180$ GeV (consistent with \cite{R17}) and $m_b=5$ GeV
for the top and bottom quark masses and a constant value for the strong
coupling constant $\alpha_s =0.12$ [the value of $\alpha_s$ in the
running from a scale of 0.1 to 1 TeV does not change significantly]; the
other fixed input parameters are $\alpha=1/128$, $M_Z=91.187$ GeV and
$s_W^2=0.23$ \cite{R18}. For the SUSY parameters, we will take into 
account the experimental bounds from the Tevatron and LEP1.5 data 
\cite{R16}, and in some cases use the values favored by fits 
of the electroweak precision data from LEP1 \cite{R15}. \s 

Fig.~2 shows the partial widths for the decays of the lightest top
squark into the two charginos $\chi_{1,2}^+$ and a bottom quark [2a] 
and into the lightest neutralino $\chi_1^0$ and the sum of all neutralinos
[the opening of the neutralino thresholds can be seen in the curves] and a
top quark [2b]. In these figures, $\tb$ is fixed to $\tb=1.6$, a value
favored by $b$--$\tau$ Yukawa coupling unification \cite{R19}. The
solid, dashed and dot--dashed curves correspond to the $(M_2, \mu$)
values [in GeV]: $(70, -500), (70, -70)$ and $(300,-70)$ in Fig.~2a
[which give approximately the same value for the lightest chargino mass,
$m_{\chi_1^+} \simeq 70$ GeV] and $(100, -500), (100, -100)$ and
$(250,-50)$ in Fig.~2b [giving an LSP mass of $m_{\chi_1^0} \sim 50$
GeV]. These values correspond to the scenarios $M_2 \ll |\mu|$, $M_2
\simeq \mu$ and $M_2 \gg |\mu|$, and have been chosen to allow for a
comparison with the numerical analysis given in \cite{R8}. The
parameters in the $\tilde{t}$ mass matrix are fixed by requiring
$m_{\tilde{t}_2} =600$ GeV and varying $M_{\tilde{t}_L}$. The mixing
angle is then completely fixed assuming $M_{\tilde{t}_R}=
M_{\tilde{t}_L}$ ($\theta_{\tilde{t}}\approx -\pi/4$ except 
for $m_{\tilde{t}_1}$ very close to $m_{\tilde{t}_2}$);
in the bottom squark sector we have 
$m_{\tilde{b}_1}= 220$ GeV, $m_{\tilde{b}_2} \sim 230$ GeV and  
$\theta_{\tilde{b}} \simeq 0$.
\s 

Fig.~3 shows the magnitude of the QCD corrections relative to the Born
width to the decays of the lightest top squark into charginos+bottom
[3a/b] and neutralinos+stop [3c/d] for the scenarios described in
Fig.~2a [for Figs.3a/b] and Fig.~2b [for Figs.3c/d]. For both the 
neutral and charged decays, 
the QCD corrections can be rather large and vary in a wide
margin: from $\sim \pm 10\%$ for light top squarks up to $\sim -40\%$
for $m_{\tilde{t}_1} \sim m_{\tilde{t}_2}$ and some $(M_2, \mu)$ values. 
\s 

The small spikes near $m_{\tilde{t}_1} \sim 425$ (530) GeV for $\chi^+ b$
$(\chi^0 t$) decays are due to thresholds in the top squark wave function
renormalization constants from the channel $\tilde{t}_1 \ra \tilde{g} t$. 
For the depicted $m_{\tilde{t}_1}$ range, this happens only for the
value $M_2 =70$ (100) GeV which leads to $m_{\tilde{g}} \simeq 3.5 M_2
\sim 245 (350)$ GeV. Note, however, that when this occurs, the channel
$\tilde{t}_1 \ra \tilde{g} t$ becomes by far the main decay mode, 
and the chargino/neutralino modes are very rare. \s 

In Fig.~4 the variation of the QCD corrections for the decay
$\tilde{t}_1 \ra b \chi_1^+$ [4a] and $\tilde{t}_1 \ra t \chi_1^0$ [4b]
is displayed as a function of the gluino mass, for two values of $\mu=-50$ 
and $-500$ GeV and $\tb=1.6$ and $20$. The top squark masses are fixed to
$m_{\tilde{t}_1} =300$ and $m_{\tilde{t}_2}=600$ GeV 
($\theta_{\tilde{t}}= -\pi/4$) and the $\tilde{b}$ 
masses are as in Fig.~2. $M_2$ and hence the chargino and neutralino
masses are fixed by $m_{\tilde{g}}$. The figure exhibits a slight 
dependence of the QCD correction on the gluino mass. For the chosen 
set of squark mass parameters, the variation of the QCD correction 
with $\mu$ is rather pronounced, while the variation with $\tb$ 
is milder. \s 

Fig.~5 shows the partial decay widths for the decays of the
lightest bottom squark [which in our convention is denoted by
$\tilde{b}_1$ and is almost left--handed] into the lightest chargino
$\chi_{1}^-$ and a top quark [5a] and into the lightest neutralino
$\chi_1^0$ and a bottom quark [5b]. As in Fig.~2, $\tb$ is fixed to
$\tb=1.6$ and $m_{\tilde{t}_1}=600$ GeV; the mass difference between the
two squarks is $\simeq 10$ GeV and we have for the mixing angle
$\theta_{\tilde{b}} \simeq 0$. The solid, dashed and dot--dashed curves
correspond to the $(M_2, \mu$) values [in GeV]: $(60, -500), (70, -60)$
and $(300,-60)$ in Fig.~5a and $(100, -500), (100, -100)$ and 
$(250,-50)$  in
Fig.~5b. The decay $\tilde{b}_1 \ra t \chi_1^-$ is by far dominant when
the channel $\tilde{b}_1 \ra \tilde{g}b$ is closed, since its decay
width is almost two orders of magnitude larger than the $\tilde{b}_1
\ra$ LSP+bottom decay width. \s

Fig.~6 presents the magnitude of the relative QCD corrections 
to the decays $\tilde{b}_1
\ra t \chi_1^-$ [6a] and $\tilde{b}_1 \ra b \chi_1^0$ [6b] as a function
of the bottom squark mass, for the same scenarios as in Fig.~5. Again, depending
on the values of $\mu, M_2$ and $m_{\tilde{b}_1}$, the QCD corrections
vary from ($\pm$) a few percent up to $-50\%$. \s 

Finally, Fig.~7 displays the QCD corrections to the decays of the SUSY
partners of massless quarks into neutralinos, $\tilde{q} \ra q \chi_0$,
as a function of the ratio $\kappa=m_{\chi}^2/ m_{\tilde{q}}^2$ for
several values of the ratio $\gamma=m_{\tilde{g}}^2 /m_{\tilde{q}}^2,
\gamma=1.2, 1.5$ and 2 [7a] and as a function of $\gamma$ for several
values of $\kappa, \kappa=0.2, 0.5$ and $0.8$ [7b]. The quark mass and
the squark mixing angle are set to zero and all squarks are taken to
be degenerate. The corrections then depend
only on the two parameters, $\kappa$ and $\gamma$ since the dependence
on the other SUSY parameters factorizes in the Born term. The QCD
corrections vary from small [most of the time negative] values for small
$\kappa$ values and small gluino masses, up to $\sim 20\%$ near
threshold. \s 

For the decays $\tilde{q}_L \ra q' \chi^\pm_j$ [the right--handed squark
does not decay into charginos], the matrix elements in the chargino
mass matrix do not factorize in the Born expressions and the QCD 
corrections further depend on the ratios $U_{j1}/V_{j1}$ through 
the contribution $F_{\tilde{g}}$. This dependence is, however, rather mild 
since first the ratio $U_{j1}/V_{j1}$ is of order unity in most of the 
relevant SUSY parameter space [in particular for $|\mu| > M_2$] and second
the contribution $F_{\tilde{g}}$ is small compared to the other contributions
for gluino masses below 1 TeV. The QCD corrections for the decays $\tilde{q}_L 
\ra q' \chi^\pm$ are thus approximately the same as in the case of the decays 
into neutralinos. \s

In conclusion: we have calculated the ${\cal O}(\alpha_s)$ QCD corrections 
to decay modes of scalar squarks into quarks plus charginos or neutralinos
in the Minimal Supersymmetric Standard Model. We have paid a particular
attention to the case of $\tilde{t}$ [and also $\tilde{b}$] squarks, where 
mixing effects are important. In the case of top squark decays, the QCD 
corrections can reach values of the order of a few ten percent depending on
the various SUSY parameters. 
They can be either positive or negative and increase logarithmically with 
the gluino mass. For the scalar partners of light quarks, the corrections 
do not exceed the level of ten to twenty percent for gluino masses less than 
1 TeV. 

\vspace*{2cm}

\nn {\bf Acknowledgements}: \s

\nn We thank Tilman Plehn and Peter Zerwas for discussions and for the 
comparison between their results Ref.~\cite{R5a} and ours, and the 
Vienna group, in particular Sabine Kraml, for discussions about Ref.~\cite{R8}.

\newpage 

\nn \subsection*{Figure Captions}

\begin{itemize}

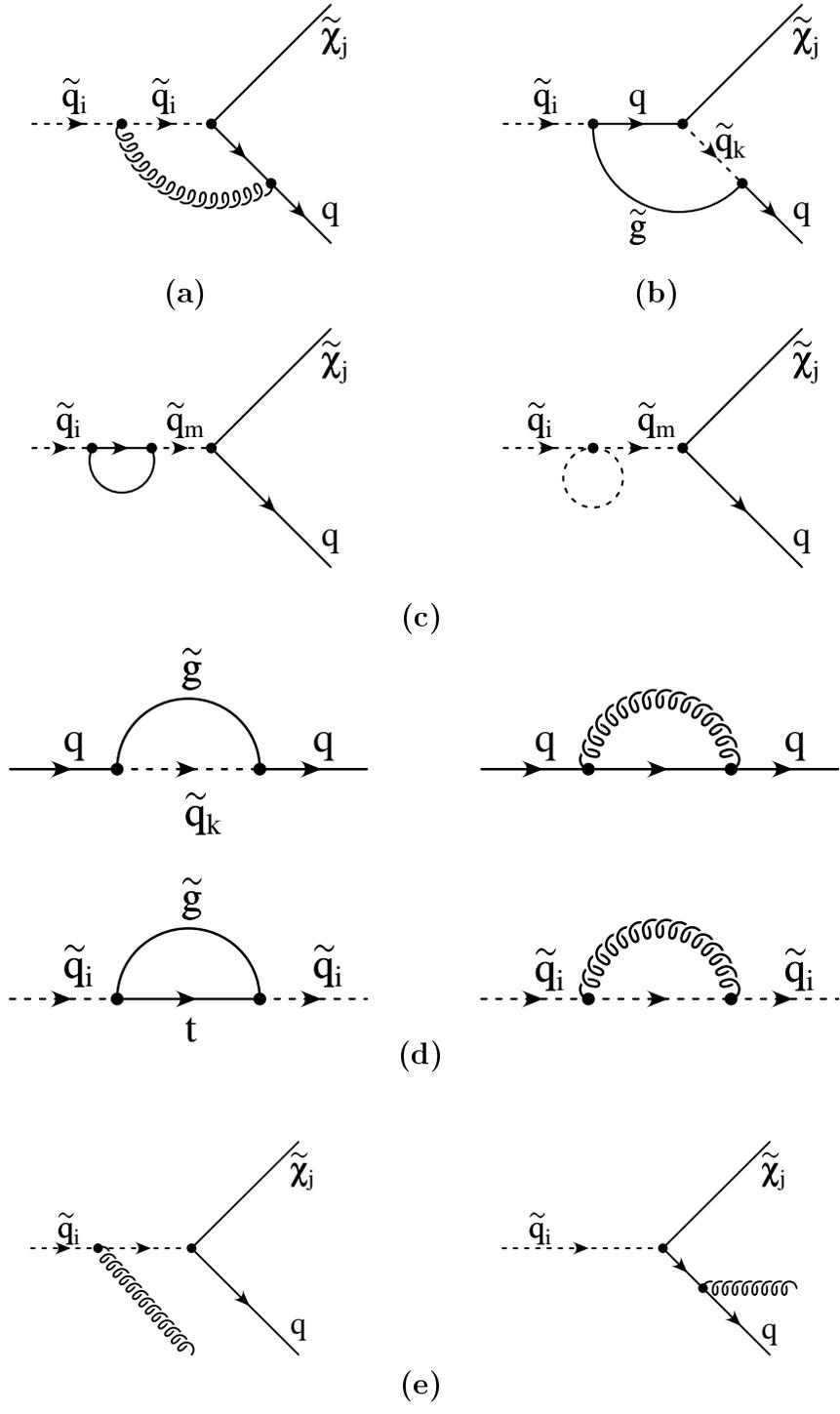
\item[{\bf Fig.~1:~}] 
Feynman diagrams relevant for the ${\cal O}(\alpha_s)$ QCD corrections 
to the decay of a squark into a quark and a neutralino or chargino. 

\item[{\bf Fig.~2:~}]
(a) Partial widths [in GeV] for the decays of the lightest top squark
$\tilde{t}_1$ into the two charginos $\chi_1^+$ and $\chi_2^+$ and a
bottom quark; $\tb=1.6$, $m_{\tilde{t}_2} =600$ GeV. The solid,
dashed and dot--dashed curves correspond to the $(M_2, \mu$) values [in
GeV]: $(70, -500)$, $(70, -70)$ and $(300,-70)$. (b) Partial widths [in GeV]
for the decays of $\tilde{t}_1$ into a top quark and the lightest neutralino 
$\chi_1^0$ as well as the sum of all neutralinos [the thresholds can be read 
off the curves]; $\tb=1.6$, $m_{\tilde{t}_2} =600$ GeV. The 
solid, dashed and dot--dashed curves correspond to the $(M_2, \mu)$
values [in GeV]: $(100, -500), (100, -100)$ and $(250,-50)$.

\item[{\bf Fig.~3:~}]
Relative size [in \%] of the ${\cal O}(\alpha_s)$ QCD corrections to the 
decay rates 
(a) $\tilde{t}_1 \ra b \chi_1^+$,  
(b) $\tilde{t}_1 \ra b \chi_2^+$, 
(c) $\tilde{t}_1 \ra t \chi_1^0$ and  
(d) $\tilde{t}_1 \ra \sum_i t \chi_i^0$, as a function of the top 
squark mass. The set of ($M_2, \mu$) parameters is as in Fig.~2a for 
Figs.~3a/b and as in Fig.~2b for Figs.~3c/d. 

\item[{\bf Fig.~4:~}]
Relative size [in \%] of the ${\cal O}(\alpha_s)$ QCD corrections to the 
decay rates (a) $\tilde{t}_1 \ra b \chi_1^+$ and (b) $\tilde{t}_1 \ra t 
\chi_1^0$ as a function of the gluino mass. $M_2$ is fixed in terms of
$m_{\tilde{g}}$ with the GUT relation. The solid, dashed, dotted and 
dot--dashed curves correspond to the $(\tb, \mu$) values [$\mu$ is in GeV]: 
(1.6, -50), (20, -50), (1.6, -500) and (20, -500).

\item[{\bf Fig.~5:~}]
(a) Partial widths [in GeV] for the decays of $\tilde{b}_1$ into the 
lightest chargino $\chi_1^+$ and a top quark; $\tb=1.6$, $m_{\tilde{t}_2} 
=600$ GeV. The solid, dashed and dot--dashed curves correspond to the 
$(M_2, \mu$) values [in GeV]: (60, -500), (70, -60) and (300,-60). 
(b) Partial widths [in GeV] for the decays of $\tilde{b}_1$ to lightest 
neutralino $\chi_1^0$ and a bottom quark; again $\tb=1.6$, $m_{\tilde{t}_2} 
=600$ GeV. The solid, dashed and dot--dashed curves correspond to the 
$(M_2, \mu$) values [in GeV]: (100, -500), (100, -100) and (250,-50). 

\item[{\bf Fig.~6:~}]
Relative size [in \%] of the ${\cal O}(\alpha_s)$ QCD corrections to the 
decay rates (a) $\tilde{b}_1 \ra t \chi_1^-$ and (b) $\tilde{b}_1 \ra b 
\chi_1^0$ and as a function of the bottom squark mass. The set of ($M_2, 
\mu$) parameters is as in Fig.~5a/b.

\item[{\bf Fig.~7:~}]
The size of the QCD corrections for the decays $\tilde{q}_{L} \ra q 
\chi^0$ as a function of the ratios $m_\chi^2 / m_{\tilde{q}}^2$ (a)
and $m_{\tilde{g}}^2/m_{\tilde{q}}^2$ (b). The solid, dashed and dot--dashed 
lines correspond respectively to $\gamma=1.2, 1.5$ and $2$ in (a)
and $\kappa=0.2,0.5$ and $0.8$ in (b). 
 
\end{itemize}

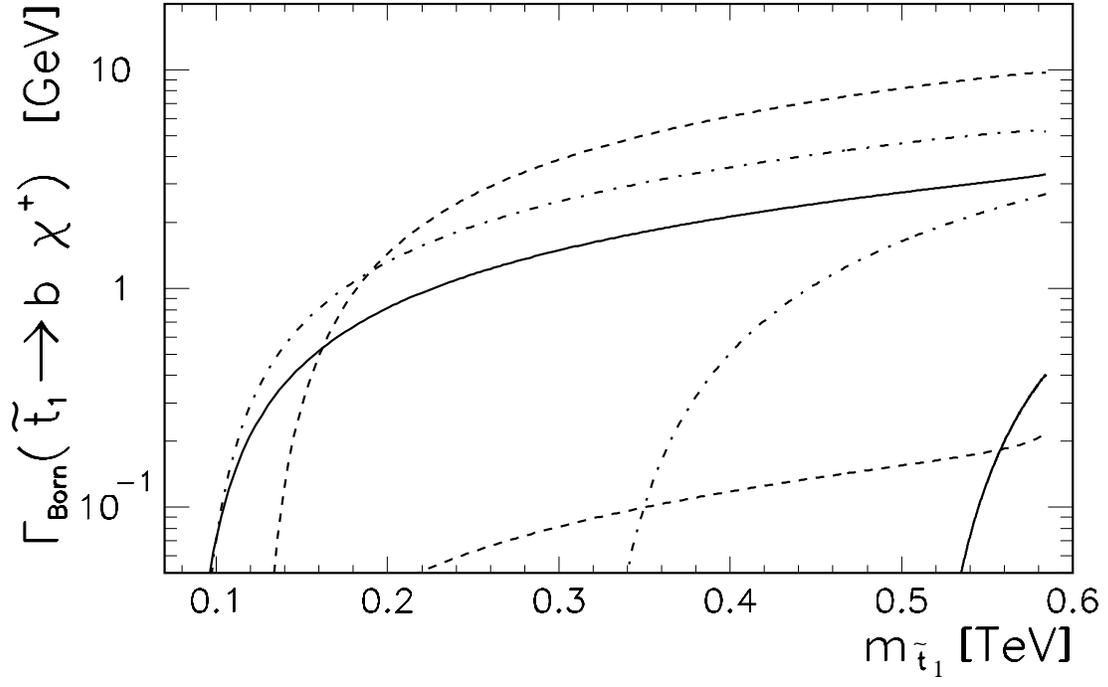
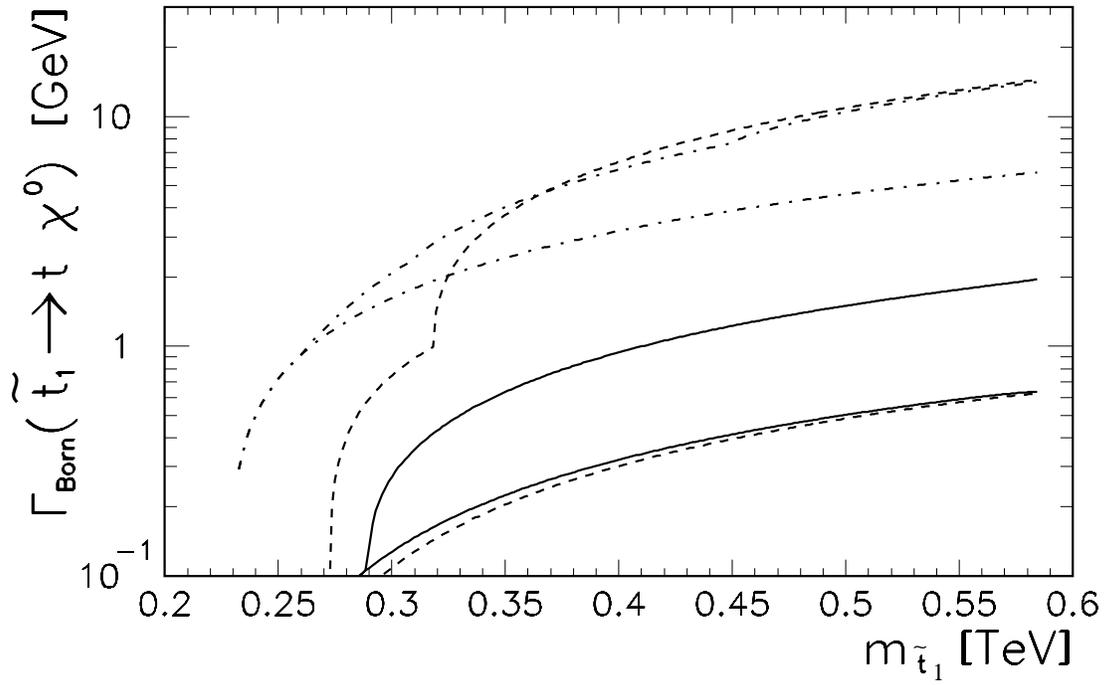
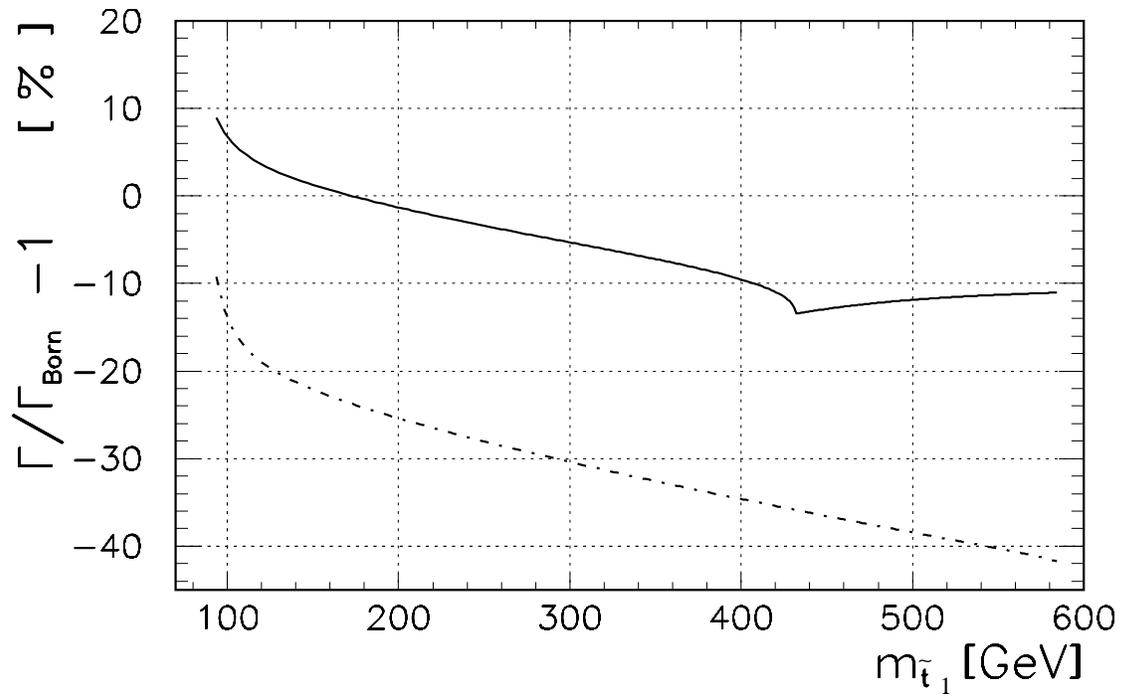
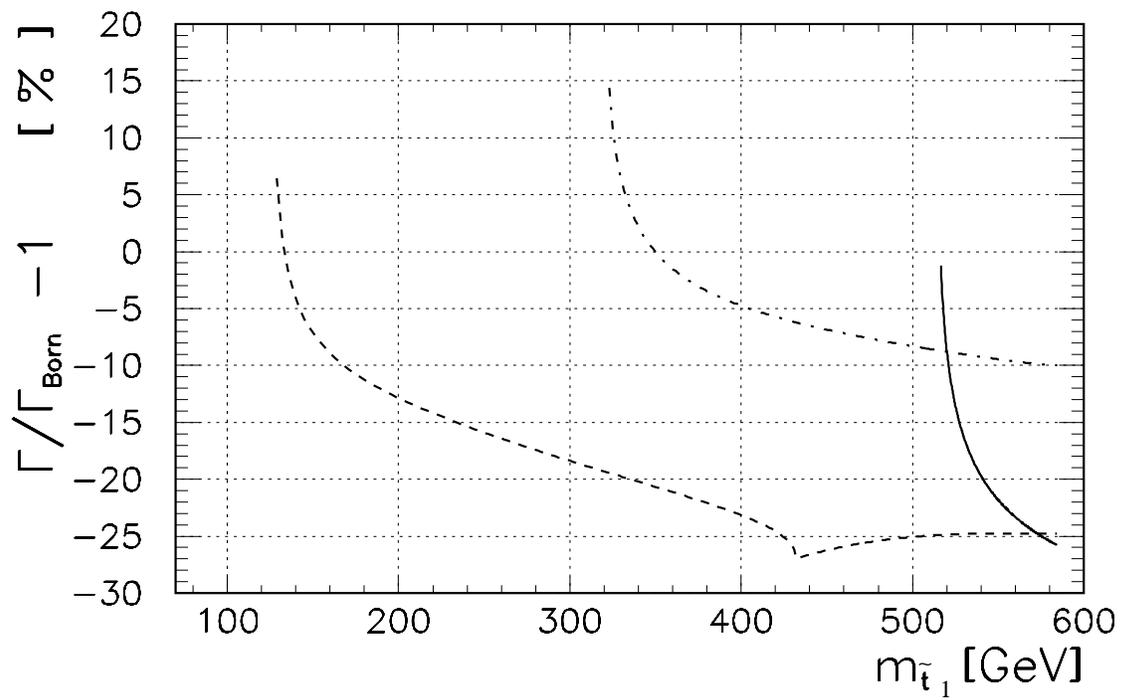
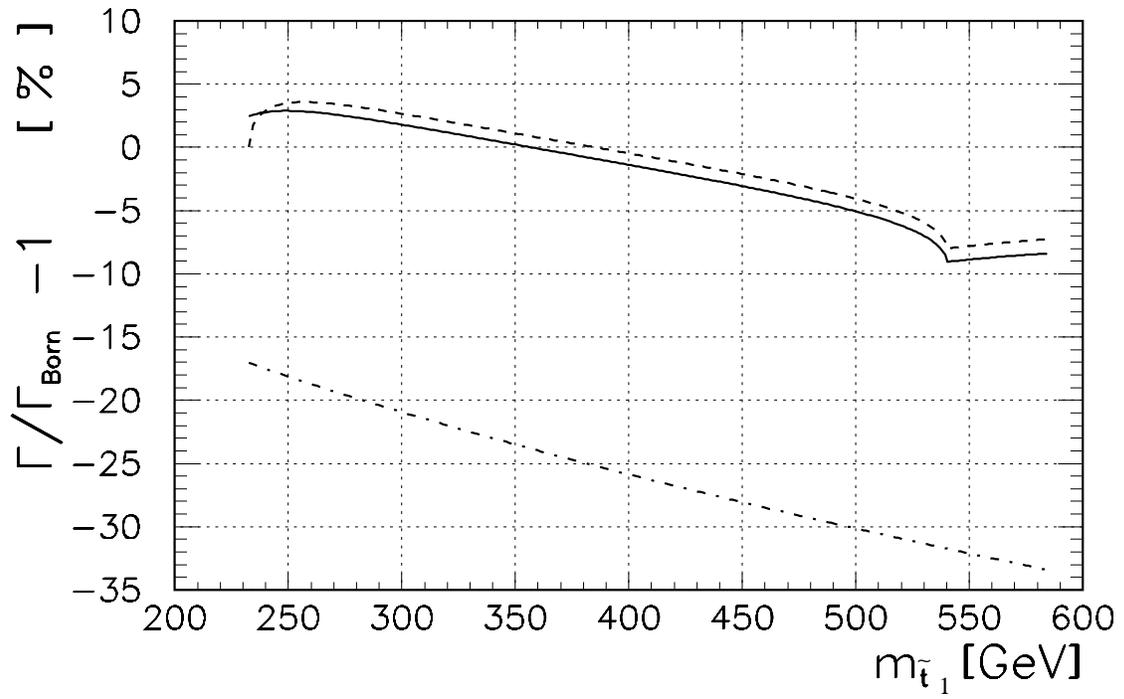
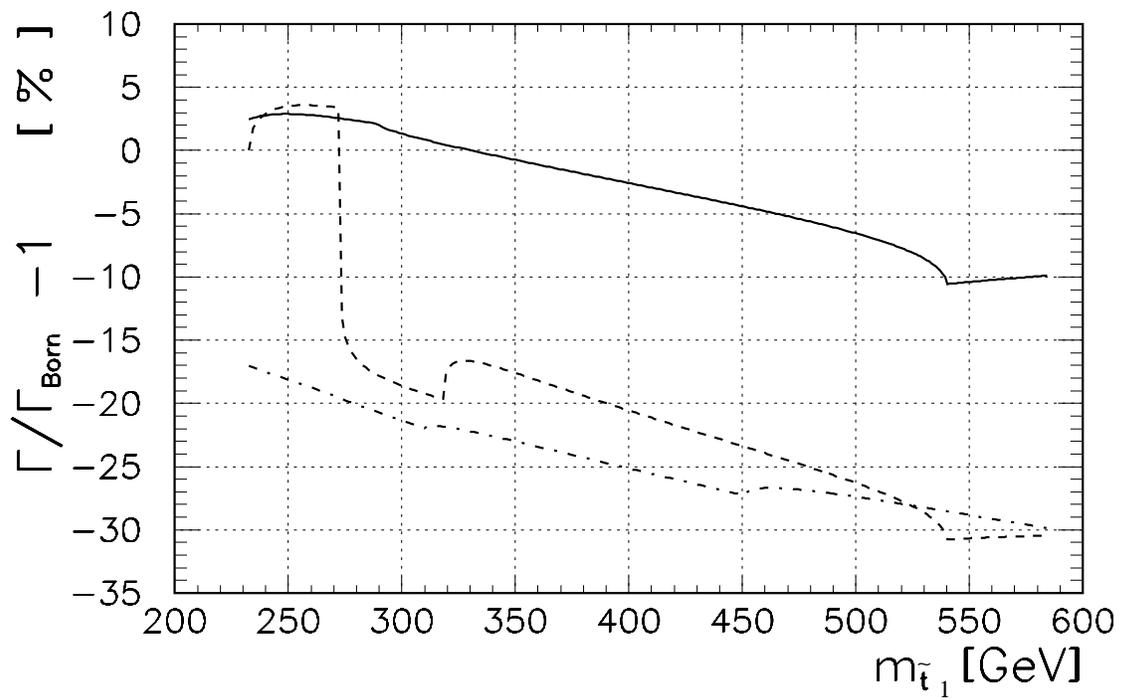
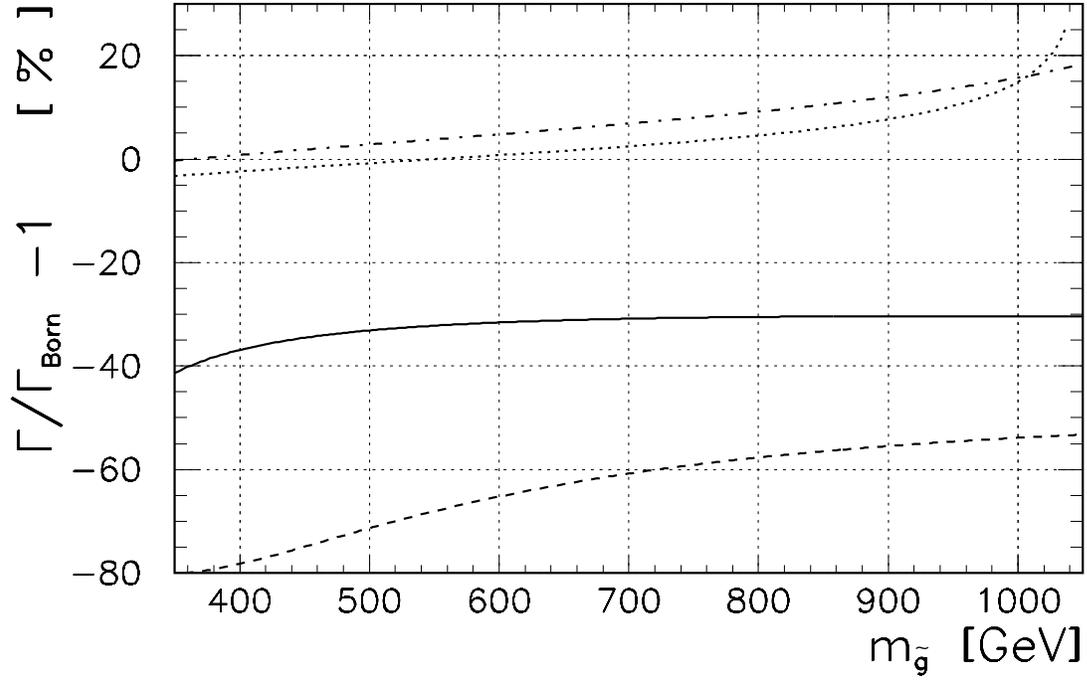
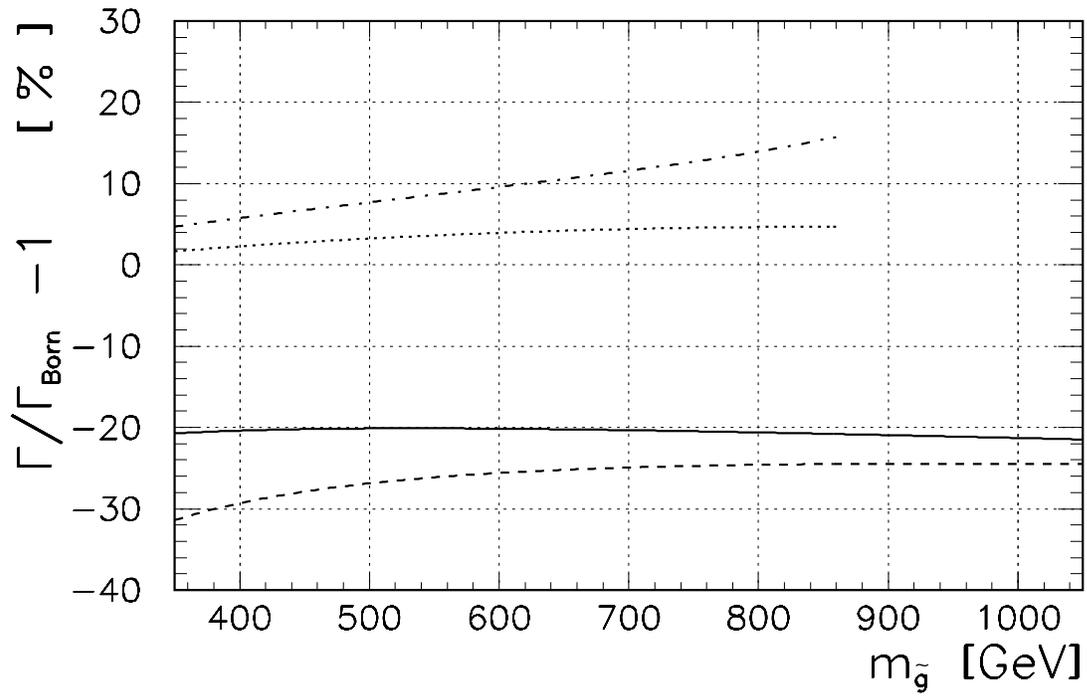
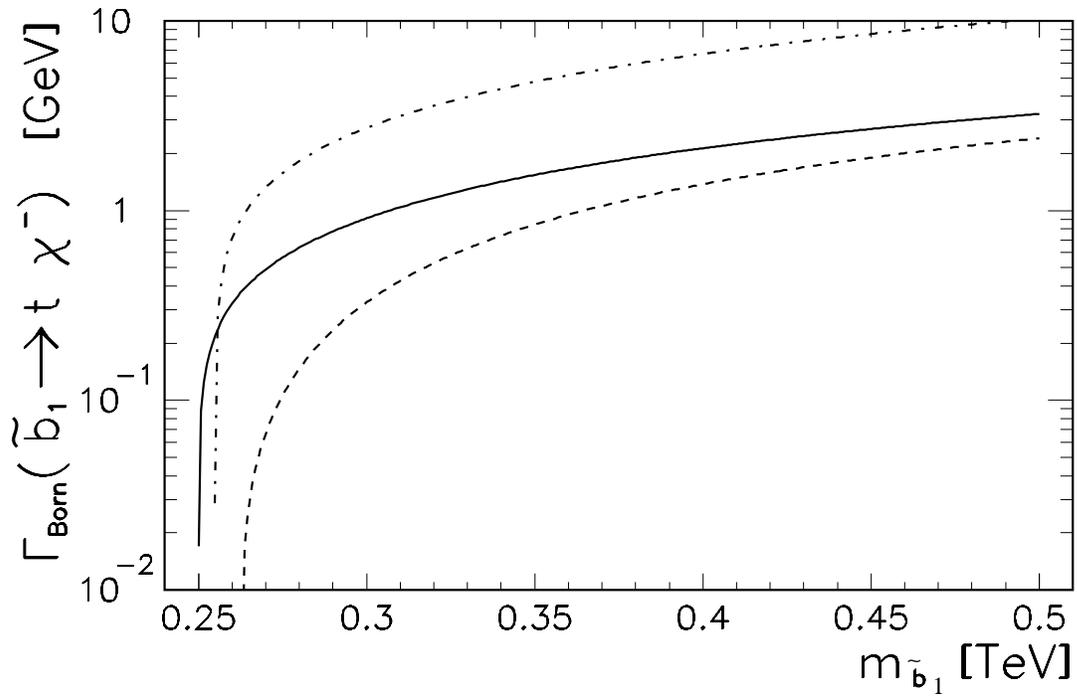
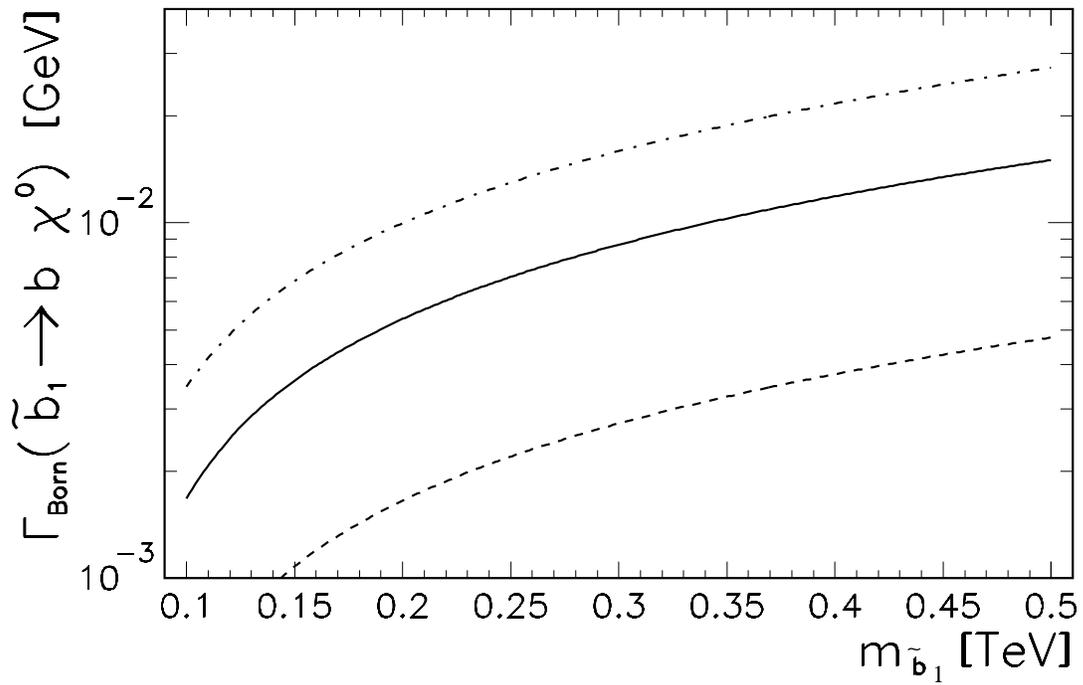
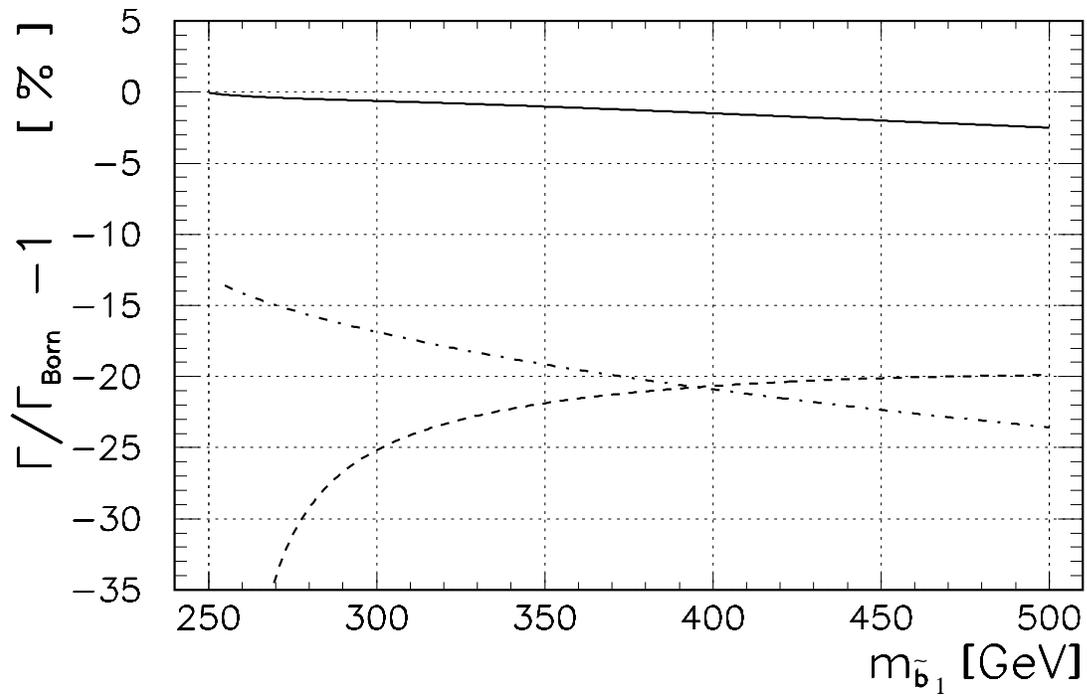
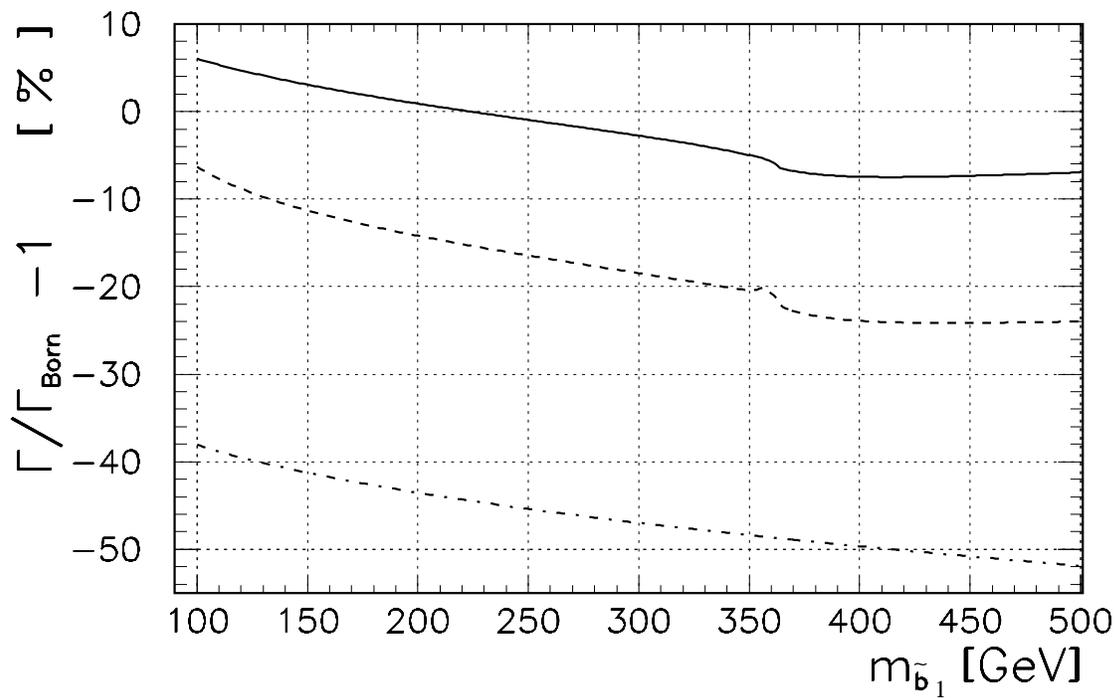
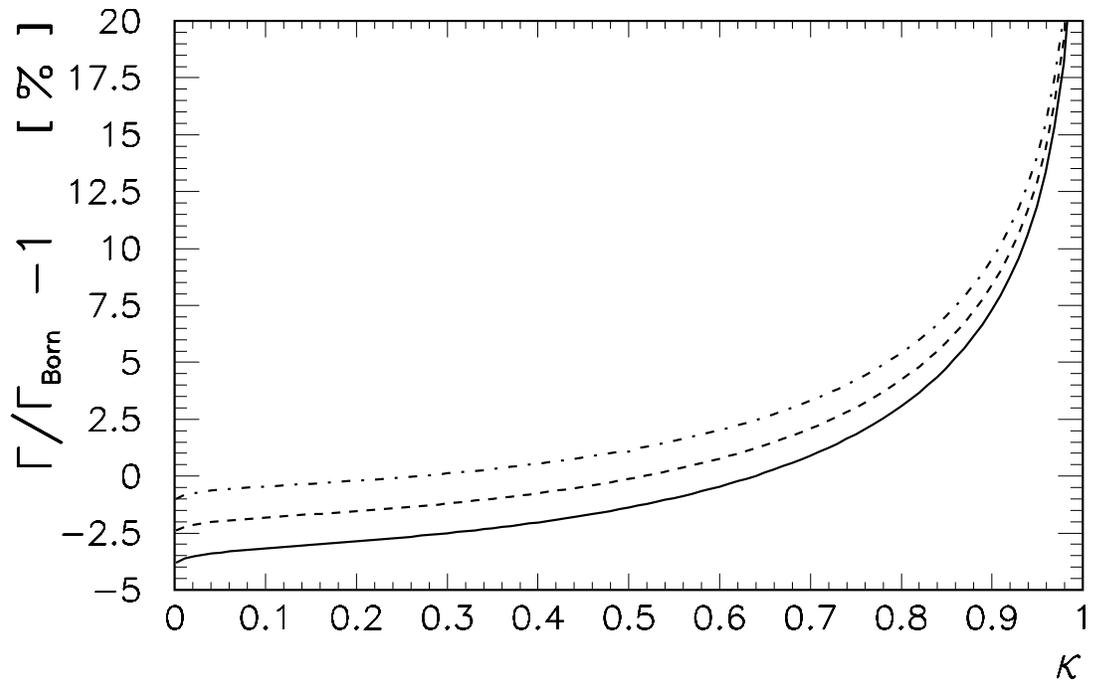
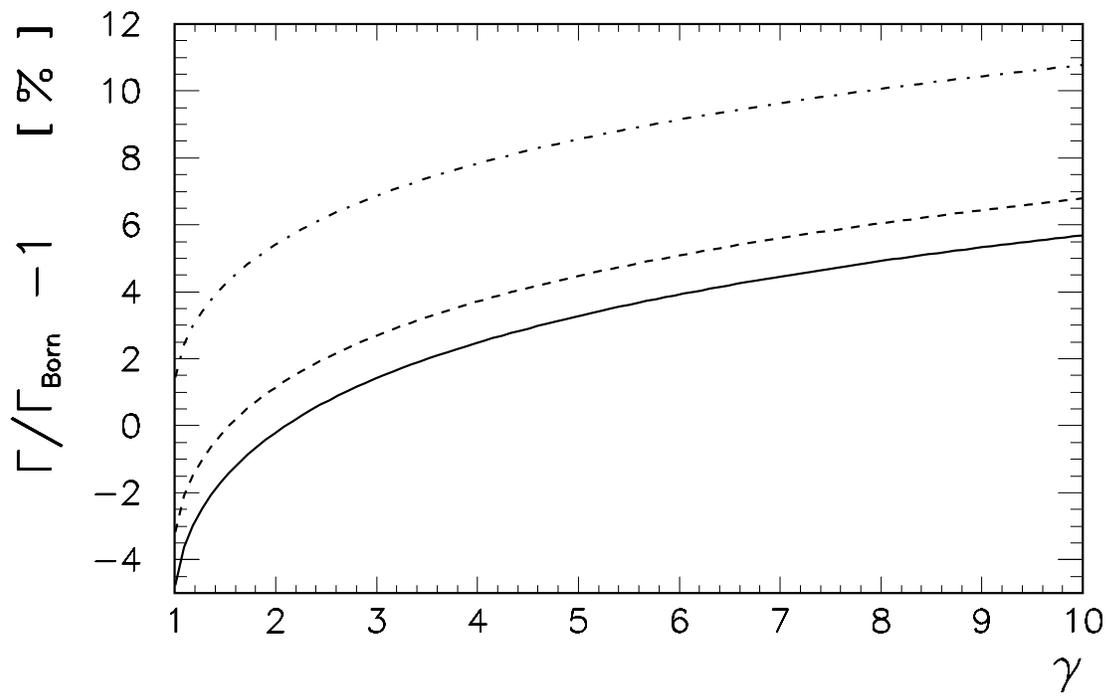
\end{document}